\documentclass[aps,prl,twocolumn,superscriptaddress]{revtex4}
\usepackage{graphicx,bm}
\begin{document}

\title{Kinematics of the swimming of \emph{Spiroplasma}}

\author{Jing Yang}
\affiliation{Center for Cell Analysis \& Modeling, University of Connecticut Health Center, Farmington, Connecticut 06030, USA}
\author{Charles W. Wolgemuth}
\affiliation{Center for Cell Analysis \& Modeling, University of Connecticut Health Center, Farmington, Connecticut 06030, USA}
\affiliation{Department of Cell Biology, University of Connecticut Health Center, Farmington, Connecticut 06030, USA}
\author{Greg Huber}
\thanks{Author to whom correspondence should be addressed (Email: huber@uchc.edu)}
\affiliation{Center for Cell Analysis \& Modeling, University of Connecticut Health Center, Farmington, Connecticut 06030, USA}
\affiliation{Department of Cell Biology, University of Connecticut Health Center, Farmington, Connecticut 06030, USA}
\affiliation{Department of Mathematics, University of Connecticut, Storrs, Connecticut 06269, USA}

\date{\today}

\begin{abstract}
\emph{Spiroplasma} swimming is studied with a simple model based on resistive-force theory. Specifically, we consider a bacterium shaped in the form of a helix that propagates traveling-wave distortions which flip the handedness of the helical cell body. We treat cell length, pitch angle, kink velocity, and distance between kinks as parameters and calculate the swimming velocity that arises due to the distortions. We find that, for a fixed pitch angle, scaling collapses the swimming velocity (and the swimming efficiency) to a universal curve that depends only on the ratio of the distance between kinks to the cell length. Simultaneously optimizing the swimming efficiency with respect to inter-kink length and pitch angle, we find that the optimal pitch angle is 35.5$^\circ$ and the optimal inter-kink length ratio is 0.338, values in good agreement with experimental observations. 
\end{abstract}


\maketitle

\begin{figure}[h]
\includegraphics[width=1.5in, angle=90]{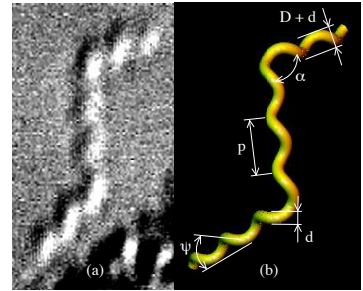}\\
\caption{\label{fig:schematic}(a) \emph{Spiroplasma} in differential interference contrast (Image courtesy of J.~W.~Shaevitz). (b) Schematic of helical geometry. $D$ is the helix diameter, $D + d$ is the outer coil diameter ($D=0.138\mu m$, $d=0.20\mu m$). Here the bend angle $\alpha$ and pitch angle $\psi$ satisfy $\alpha = \pi-2\psi$.}
\end{figure}

Helices are profoundly involved in the swimming of many types of bacteria. For example, the classical picture of bacterial swimming involves long helical filaments extending out from the inner membrane of the cell~\cite{berg_93}, driving the bacterium forward by their rotation. But helices are also used in other more ingenious ways. In \emph{Treponema pallidum}, the spirochetal bacterium responsible for syphilis, rotation of helical filaments encased between the bacterial cell wall and the outer membrane leads to periodic undulations that drive this cell through water and other viscous fluids. (This is not, presumably, how it crossed the Atlantic Ocean \cite{Fracastoro_1530}.) The \emph{Leptospiraceae} rely on a helical body plan and rotation of tightly-coiled filaments to drive their motility~\cite{berg_1978}. Generally speaking, it is the dynamics and shape together, the ``swimming strategy'', that is the key to understanding the motility of helical bacteria. For the flagella-less family of plant pathogens represented by \emph{Spiroplasma}, the precise propulsive strategy was shrouded in mystery until careful observations \cite{gilad_2003, shaevitz_2005} revealed that these tiny organisms processively flip the chirality of regions of their helical bodies, generating pairs of kinks moving down the cell body (Fig. \ref{fig:schematic}(a)). The internal propulsive mechanism underlying this peculiar swimming strategy is still unclear, as is the quantification of its motility.

\emph{Spiroplasma} swimming is realized by the propagation of a pair of kinks along the body axis of the cell, and the bacterium is propelled by the hydrodynamic force as the fluid associated with the kinks moves rearward with the ``body wave''. Kinks start at the same end of the cell (front), and travel toward the other end (back). As the kink propagates, the cell changes direction through an angle related to the pitch angle of the helix. Because of the unbalanced viscous drag between different portions of the cell body separated by kinks, eventually the cell swims in a zigzag path. The helical transformation requires a cell to rotate about its body axis during kink propagation \cite{goldstein_2000, wolgemuth_2003, shaevitz_2005}.

In this Letter we use the experimentally-observed geometry of \emph{Spiroplasma} to explore its swimming behavior for a range of kinematic deformations. We consider a helically-shaped cell of fixed pitch angle and helix diameter. To swim, the bacterium flips the handedness of its helix beginning at one end. This change in helicity propagates toward the back of the cell with a uniform velocity, $U_k=10.5\pm0.3$ $\mu$m/s \cite{shaevitz_2005}. After a time $t_k$, the front of the cell reverts to its original handedness and this change in helicity also propagates toward the back of the cell with the same kink velocity. The distance between the kinks is $L_k = U_k t_k$, and only an integer number of turns is considered. We treat $U_k$, $L_k$, and the cell length, $L_c$, as parameters and use resistive-force theory (RFT) to calculate the swimming speed and trajectory as a function of these parameters. Our model is complementary to the work by Wada and Netz \cite{wada_2007} where slender-body hydrodynamics \cite{rotne_1969} was used to explore the dependence of swimming speed on the pitch angle, $\psi$. By varying the distance between kinks, we examine both the geometry and kinematics of the swimming process. It was found experimentally that the kink velocity $U_{k}$ \cite{gilad_2003, shaevitz_2005} and the inter-kink distance $L_{k}$ \cite{shaevitz_2005} show a remarkable consistency -- what determines this preference? Our results demonstrate that the maximum propulsive efficiency is achieved when the pitch angle is around 35.5$^{\circ}$ (in agreement with \cite{wada_2007}) and the ratio between inter-kink length and cell-body length is around 0.338, which is consistent with most experimental observations and suggests that the kinematics of \emph{Spiroplasma} has been evolutionarily tuned.

Any helix can be described by its pitch, $p$, and helix radius, $r$. One helical repeat in the arclength $s$ is $\ell=\sqrt{p^{2}+4 \pi^{2} r^{2}}$. When two helices of differing pitch are concatenated, such that the tangent and normal vectors are continuous, a kink is formed \cite{hotani_1976, goldstein_2000}. The bend angle $\alpha$ between the two axes in this construction is $\alpha = \pi - 2 \psi$ (Fig.~\ref{fig:schematic}(b)). The parameters employed in the calculations are $p=0.62$ $\mu$m, $r=D/2=0.069$ $\mu$m, and $\psi=35^{\circ}$ \cite{gilad_2003, shaevitz_2005}. For the numerical integration, the cell body is discretized into straight-line elements. According to RFT \cite{gray_1955, becker_2003}, the resistance force acting on a unit length of rod is given by ${\mathbf{f}} = - \left[ 2\pi\mu/\ln(2/\epsilon) \right] \left( 2 {\mathbf{I}}-{\boldsymbol{\lambda \lambda}} \right) \cdot {\mathbf{U}}+{\boldsymbol{O}} \left[ \ln ^{-2} (2/\epsilon) \right]$, where ${\mathbf{U}}$ is the local velocity, ${\mathbf{I}}$ is the unit tensor, $\epsilon$ is the slenderness ratio of rod diameter to length, ${\boldsymbol{\lambda}}$ is the unit tangent vector, and $\mu$ is viscosity. The total force and momentum on the unrestrained cell body is zero, i.e., $\int^{L_{c}}_{0} {\mathbf{f}}(s) ds = {\boldsymbol{0}}$, $\int^{L_{c}}_{0} {\mathbf{M}}(s) ds = {\boldsymbol{0}}$. The translational and rotational velocities of the cell body can be solved for from the resulting system of linear equations. To verify that the implementation of the model has been properly made, the translational velocity of a rigid helix with angular velocity $\omega$ around its body axis is calculated with the same numerical code. Since the total force on the rotating helix is zero, i.e., $\int^{L}_{0}{\mathbf{f}}(s)ds={\boldsymbol{0}}$ ($L$ is the total length of the rigid helix), the translational velocity of the helix is $ U= B \ell \omega $, where $ B = \sin^{2} \psi \cos \psi / \left[ 2 \pi ( 1 + \sin^{2} \psi ) \right] $ \cite{wada_2007}. The numerical results match this resistive-force solution with relative errors less than 0.05\% under our typical spatial discretization. A slender-body theory (SBT) calculation, using the straight line elements, was also performed to account for hydrodynamic interactions. Based on the aforementioned model, a number of dynamical questions on the swimming of \emph{Spiroplasma} can be formulated and addressed.

\begin{figure}[h]
\includegraphics[width=2.3in]{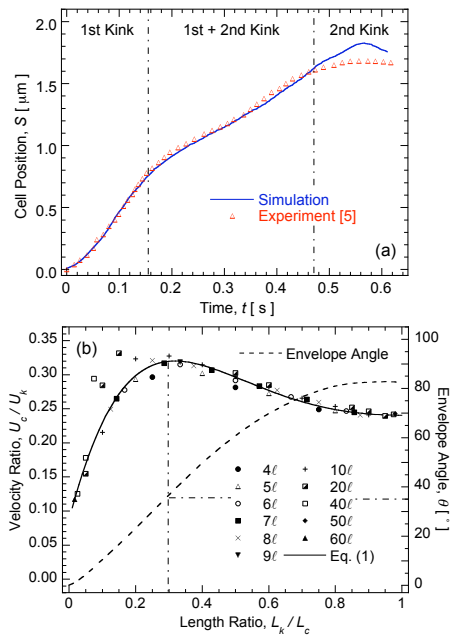}
\caption{ \label{fig:displacement} (a) A simulated displacement curve ($L_{k}$=2$\ell$, $L_{c}$=7$\ell$, and $U_{k}$=12$\ell$/s). (b) Velocity ratio and the corresponding envelope angle, as a function of the length ratio for different body lengths, from RFT (fit by solid line); Maximum velocity ratio is located where envelope and pitch angle are equal.}
\end{figure}

Typical experimental and simulated displacement curves during one swimming cycle of \emph{Spiroplasma} are shown in Fig.~\ref{fig:displacement}(a). Here the displacement is defined as the projected center-of-mass position. For comparison with the experimental data, our numerical simulation results are also smoothed with a 175-ms boxcar filter \cite{shaevitz_2005, boxcar}. Using the parameters given above, our simulations agree well with the corresponding experimental results \cite{brownian_motion}, giving an average velocity of about 3 $\mu$m/s, close to the observed cell velocity, 3.3$\pm$0.2 $\mu$m/s \cite{shaevitz_2005}.

It is observed experimentally that the speed is decreased when two kinks are present as compared with when only one kink is present. This key feature of the displacement curve is reproduced in the numerical simulations (Fig.~\ref{fig:displacement}(a)). Our simulations show that the cell velocity increases linearly with kink velocity, and that the ratio of cell velocity to kink velocity is constant for fixed inter-kink length, with the proportionality constant varying with inter-kink length, consistent with the experimental reports \cite{gilad_2003, shaevitz_2005}.

\begin{figure}[h]
\includegraphics[width=2.0in,angle=-90]{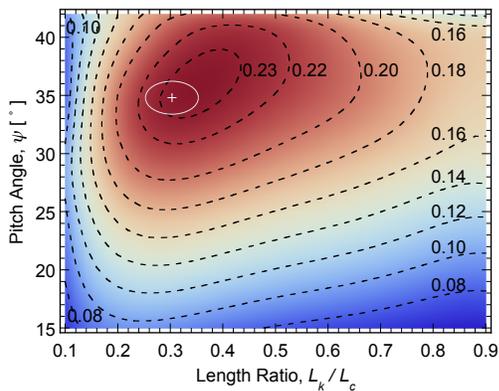}
\caption{\label{fig:opt}Velocity ratio as a function of kink length ratio and pitch angle. Contour lines are velocity-ratio isolines. Experimental observations are centered within the error ellipse.}
\end{figure}

In order to delineate the effect of dynamically changing geometry on \emph{Spiroplasma} motility, two dimensionless parameters are introduced. One is the ratio of cell velocity $U_{c}$ to kink velocity $U_{k}$; the other is the ratio of inter-kink length $L_{k}$ to cell body length $L_{c}$. We find that all the data fall on a single curve after the results from different cell-body lengths and inter-kink lengths are taken together (Fig.~\ref{fig:displacement}(b)). The curve describes a velocity ratio that increases with length ratio at small values, and then begins to decrease with length ratio after reaching a peak value. From Fig.~\ref{fig:displacement}(b), the maximum velocity ratio of the data-fitted curve is around 0.323 when the length ratio is about 0.32. The fact that the velocity ratios and length ratios observed in the experiment fall close to the aforementioned value suggests that wildtype \emph{Spiroplasmas} have optimized their geometry to achieve the ``fastest state'' in the course of evolution. The study of Wada and Netz \cite{wada_2007} suggests an optimized pitch angle as well, the treatment of which goes beyond the reach of RFT for pitch angles greater than roughly 35$^{\circ}$ \cite{jung_2007,jung_2007_2}. The full optimization over both length ratio and pitch angle (Fig.~\ref{fig:opt}) using our SBT code gives a maximum velocity ratio at $L_{k}/L_{c}=0.338$ and $\psi=35.5^{\circ}$. These values are consistent with the experimental values 0.303$\pm0.051$ and 34.8$\pm1.4^\circ$ respectively \cite{shaevitz_2005}.

The velocity ratio can be calculated in RFT in the limit that the length ratio approaches one. As the length ratio goes to one, there is effectively only one kink propagating along the cell body at any given time. In this case, \emph{Spiroplasma} can be treated as a filament consisting of two helical sections, one right handed and the other left handed. The two helical sections rotate around their respective axes with the opposite sense. The total translational velocity is the vector sum of these two components with an interior angle $2 \psi$. From geometrical considerations, the kink velocity $U_{k}$ can be represented as a function of rotational velocity $\omega$, i.e., $U_{k} = 2 \omega r / \tan \psi$. Using the previously-mentioned translation velocity of a rigid helix ($B\ell\omega$) the total velocity can be estimated to be $2 B \ell \omega \cos \psi$, and the corresponding velocity ratio reduces to $2 \pi B = \sin^{2}\psi \cos \psi / \left(1+\sin^{2}\psi \right)$. For a pitch angle of $35^{\circ}$, this ratio is around 0.202, not far from the simulation result of 0.24. At a length ratio of one, the kinematic asymmetry between the left-handed inter-kink length and the right-handed extra-kink length vanishes, and a chirality variable $\chi\equiv(L_{c}-L_{k})/(L_{c}+L_{k})$ can be used to study the velocity ratio on both sides of $\chi=0$ \cite{datfitexp}. We note in passing that the entire velocity-ratio curve is in excellent empirical agreement with a quartic polynomial \cite{sbt_curve}
\begin{equation}\label{datafit}
U_{c}/U_{k} \cong 0.24+0.73\phi^{2}-1.61\phi^{4}
\end{equation}
\*(the solid curve in Fig.~\ref{fig:displacement} (b)) where $\phi=\tan^{-1}\chi$. 

\begin{figure}[h]
\includegraphics[width=1.2in,angle=-90]{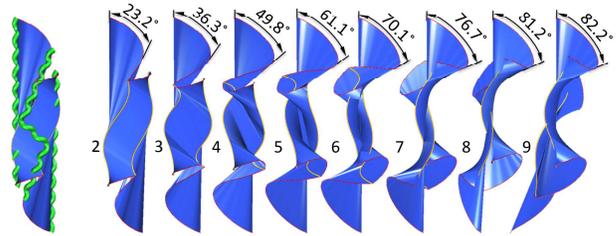}
\caption{\label{fig:Env_Angle}The trace swept out by the cell-body axis at eight different inter-kink lengths 2-9 $\ell$ (for cell length 10 $\ell$). Three snapshots of the cell body are superposed in the leftmost trace ($L_{k}=2 \ell$). Yellow lines indicate the trajectories of the two kinks. The envelope angle is indicated.}
\end{figure}

Not only translation, but also rotation of the cell body must satisfy the dynamic balance, i.e., zero force and torque. At the beginning of a kink cycle, one end of the cell-body axis traces out a circular arc (see Fig.~\ref{fig:Env_Angle}). The maximum angle swept out, we term the \emph{envelope angle}. We use this concept in Fig.~\ref{fig:displacement}(b), where the dashed curve shows that the envelope angle increases with length ratio. Fig.~\ref{fig:displacement}(b) also shows that the velocity ratio achieves its maximum at the length ratio whose envelope angle equals the pitch angle (see the dot-dashed line).

\begin{figure}[h]
\includegraphics[width=2.0in,angle=-90]{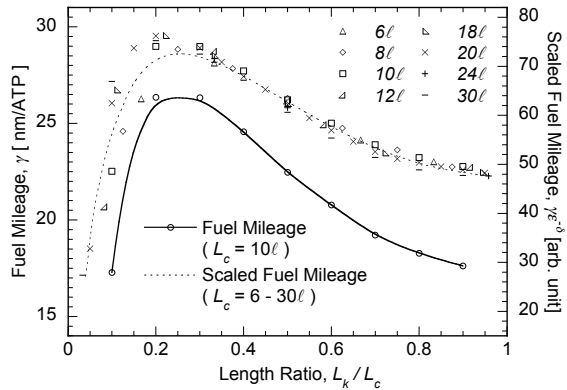}
\caption{\label{fig:optimiz} Fuel mileage as a function of the length ratio. The solid curve and circle data points show the case with $L_{c}=10 \ell$ (left ordinate). Using the scaling factor $\epsilon^{-\delta}$, the results from all cell lengths can be collapsed. Here $\epsilon$ is the slenderness ratio and the exponent $\delta = 1.90 \pm 0.01$ \cite{fig4com}.}
\end{figure}

Experiments on \emph{Spiroplasma} do not observe more than two kinks along the cell body at one time. It is therefore interesting to consider the swimming efficiency as a function of the number of kinks per cycle in order to determine whether a two-kink cycle is optimal. A standard definition of swimming efficiency compares the energy dissipated by dragging the appropriate rigid cylinder through fluid to the power dissipated by the actual swimmer \cite{wada_2007,becker_2003,tam_2007}. This definition compares a swimming object to a non-swimmer and also does not account for the energy dissipated to create the swimming motion. Since an unknown mechanism is responsible for producing the traveling kinks in \emph{Spiroplasma}, we assume that there is a fixed net energy, $\delta E$, required to produce a \emph{pair} of kinks \cite{morekinks} and that this mechanism sets a uniform kink velocity. Under these assumptions, we hypothesize that biology tries to maximize the distance traveled per unit energy expended. Therefore, we consider the \emph{fuel mileage} $\gamma$ = distance traveled per energy consumed ($n \delta E$ + the energy dissipated by the fluid, where $n$ is the number of kink pairs per cycle), which we compute in terms of nanometers traveled per ATP burned \cite{mpg}. First, ignoring the energy to create the kinks ($\delta E\simeq0$), we calculate $\gamma$ as a function of the length ratio (solid curve in Fig.~\ref{fig:optimiz}). The fuel mileage from different cell lengths can all be collapsed onto one curve by constructing a scaling factor from a power of the slenderness ratio $\epsilon$ ($=d/L_{c}$) as $\epsilon^{-\delta}\gamma(L_{k},L_{c})$ with $\delta \simeq 2$. In both RFT and SBT, we find that the fuel mileage is maximized at a length ratio of around 0.3, although the methods’ different velocities lead to shifted $\gamma$ values: $\gamma_{_{\mathrm{RFT}}}/\gamma_{_{\mathrm{SBT}}}\cong3-3.4$. Using RFT, we find that a single double kink cycle requires 50-100 ATPs worth of energy and the cell moves about 26 nm/ATP, whereas SBT gives 150-340 ATPs per cycle and the cell moves 8 nm/ATP. Interestingly, the latter is the fuel mileage of a single kinesin molecule \cite{schnitzer_1997}. When $\delta E$ is small, we find that the fuel mileage for a 4-kink cycle is larger than for a 2-kink cycle. However, when $\delta E$ is greater than 40 ATPs, then a 2-kink cycle is optimal. Therefore, we expect that the mechanism producing the double kink must require at least 40 ATPs per cycle.

In this Letter, we investigated via a numerical study the swimming of \emph{Spiroplasma} with a model based on resistive-force theory and the kinematic deformations of \emph{Spiroplasma}. The key features of \emph{Spiroplasma} swimming are well reproduced with this model, and the simulation results agree quantitatively with the observed experimental data. The important effects of dynamic geometry on the motility of \emph{Spiroplamsa} are explored, and the motility efficiency is shown to assume the highest possible value given the kinematics observed in experiments. One implication is the optimization of the geometry and kinematics of \emph{Spiroplasma} during the course of evolution.

\begin{acknowledgments}
We gratefully acknowledge National Institute of Health grants R01 GM072004 (CWW) and U54 RR022232 (CWW, GH), and support from the Richard Berlin Center for Cell Analysis \& Modeling (JY, GH).\end{acknowledgments}


\begin{thebibliography}{99}
\bibitem[{box()}]{boxcar}
\eprint{The boxcar filter uses $S_{BC}(t)=\frac{1}{2 \sigma} \int^{t+\sigma}_{t-\sigma}S\left(t^{\prime}\right)dt^{\prime}$, where $S$ is the displacement before filtering, $S_{BC}$ is the displacement after filtering, and $\sigma$ is the half width.}
\bibitem[{sbtcurve()}]{sbt_curve}
\eprint{The corresponding velocity ratio from SBT has a similar fit $U_{c}/U_{k} \cong 0.165+0.73\phi^{2}-1.82\phi^{4}$, to leading order a shift of 0.075 from Eq. (\ref{datafit}).}
\bibitem[{morek()}]{morekinks}
\eprint{Because a kink flips the sign of the pitch, a complete cycle must involve an even number of kinks, in order to return the system to its original state at the end of the cycle.}
\bibitem[{atp2mpg()}]{mpg}
\eprint{Given that 1 ATP provides 20 k$_{\mathrm{B}}$T, then 1 nm/ATP = (83 pN)$^{-1}$. This is very nearly 1 mpfg (miles per femtogallon of gasoline).}
\end{thebibliography}
\end{document}